# A logical representation of Arabic questions toward automatic passage extraction from the Web

**Wided BAKARI**
Faculty of Economics and Management
3018, Sfax Tunisia
MIR@CL, Sfax, Tunisia
Wided.bakkari@fsegs.rnu.tn

**Patrice BELLOT**
Aix-Marseille University, F-13397
LSIS, Marseille, France
Patrice.bellot@gmail.com

**Mahmoud NEJI**
Faculty of Economics and Management
3018, Sfax Tunisia
MIR@CL, Sfax, Tunisia
Mahmoud.neji@gmail.com

**Abstract:** With the expanding growth of Arabic electronic data on the web, extracting information, which is actually one of the major challenges of the question-answering, is essentially used for building corpus of documents. In fact, building a corpus is a research topic that is currently referred to among some other major themes of conferences, in Natural Language Processing (NLP), such as, Information Retrieval (IR), Question-Answering (QA), Automatic Summary (AS), etc. Generally, a question-answering system provides various passages to answer the user questions. To make these passages truly informative, this system needs access to an underlying knowledge base; this requires the construction of a corpus. The aim of our research is to build an Arabic question-answering system. In addition, analyzing the question must be the first step. Next, it is essential to retrieve a passage from the web that can serve as an appropriate answer. In this paper, we propose a method to analysis the question and retrieve the passage answer in the Arabic language. For the question analysis, five factual question types are processed. Additionally, our purpose is to experiment with the generation of a logic representation from the declarative form of each question. Several studies, deal with the logic approaches in question-answering, are discussed in other languages than the Arabic language. This representation is very promising because it helps us later in the selection of a justifiable answer. The accuracy of questions that are correctly analyzed and translated into the logic form achieved 64%. And then, the results of passages of texts that are automatically generated achieved an 87% score for accuracy and a 98% score for c@1.

**Keywords:** Arabic, question analysis, answer passage retrieval, logic representation, Word Wide Web.

## 1 Introduction

Question analysis and answer passage retrieval are two tasks that are functionally dependent. Indeed, question-answering has became now one of the most popular research areas in the search for accurate information. Thus, with the evolution of digital information on the web, demands for tools that are capable of retrieving this information have also increased. This paper describes our search for tools and systems that could to be useful to satisfy users needs. In most question-answering systems, generating an accurate answer to a question in natural language necessarily involves an analysis of the question. In addition, the question analysis is an important task which is necessary not only for searching for documents but also for the extraction of a justifiable and accurate answer to it.

First, one of the principal challenges of question-answering systems is the question analysis. In our case, we proposed two sub-tasks related to the question analysis. The first is for question pre-processing that begins with a factual question and attempts to determine some elements (keyword, focus and expected answer type); these elements are later used by other modules for generating an accurate answer. The second task is for the question transformation which allows building a transformation of this question in declarative form that's will be used shortly for generating logical forms.

In recent years, many researchers have been conducted on the task of corpus construction. The majority of these investigations are based on statistical approaches. Indeed, corpus construction is a complex task because it relies in large part on a significant number of resources to be exploited; it is both delicate and essential. Thus, one advantage of a corpus is that it can easily provide quantitative data that can't provide reliably insights. In addition, corpus construction is generally used for many NLP applications, including question-answering, machine translation, information retrieval, etc. Several attempts have succeeded on building corpus. Furthermore, much research is available for the construction of corpus of documents in English and other languages. However, there are some corpus available to the public, especially in Arabic. Indeed, a corpus is a collection of pieces of texts in electronic form, selected through external criteria in order to represent a language as a data source for linguistic studies [1]. Indeed, a definition that is both specific and generic of a corpus according to Rastier [2], is the result of choices that linguists bring. A corpus is not a simple object; it should not be a "bag of words" or a mere collection of phrases. This is in fact a text assembly that can cover many types of texts.

The research in question-answering is increasing with many approaches and methodologies that are proposed for many world languages; Bulgarian, Dutch, English, Finnish, French, German, Indonesian, Italian, Japanese, Portuguese and Spanish. Few studies have been proposed for Arabic in this area. Most of these investigations deal with morphosyntactic approaches in which sophisticated linguistic analysis and natural language methods have been used. These approaches provide answers in the form of short passages, extracted from the document collection, rather than giving short and precise answers. Hence, the performance of these systems is limited by the difficulty of the Arabic language processing and the considerable lack of effective NLP tools that support Arabic [3], [4].

These days, the Web plays a very important role in the search for information. It is considered the greatest resource of knowledge (textual, graphic or sound). This source is combined with storage media that allows the rapid construction of a corpus [5]. Nevertheless, building a corpus of texts from the web was not a simple task. Such a task has contributed to the development and the improvement of several linguistic tools such as question-answering systems. The last few years have taken work to exploit this kind of data. In the framework of an automated translation in [6], others study the possibility of using the websites, which offer information in multiple languages to build a bilingual parallel corpus.

Thus, note that the Arabic language is different from the English and other languages in the order of the words, the criteria of the verb and the name. Also, it is different in the type of the sentence being treated. Nowadays, it is difficult to find a corpus designed for the processing of natural language and more specifically for the Arabic language. As well as, there have been a number of studies invested in the Arabic corpus construction, especially in Europe. However, the progress in this area is still limited. In our work, we have created a new Arabic corpus to answer questions by querying the search engine Google. We have chosen to use the Web as a source of texts because it is essentially a huge database mainly of textual documents and offers great possibilities for building corpus. We are giving a question, analyzing this question and finding the text answer, analyzing this text in order to select an answer to this question.

Thus, to consolidate our logical approach dedicated to the Arabic question-answering, presented in [7], we will proceed with a detailed analysis of the question to determine what it requests for and how to address the best answer. Our approach is based on the idea of generating logical predicates of both the questions and extracts passages containing the justifiable answer.

And yet, research on Arabic question-answering which is specifically directed to the information needs remains a little explored area. While several approaches have exploited semantic knowledge in the question-answering process, especially in English, few approaches have explored the utility of semantic logic representations of inference mechanisms. In addition, logic is a level between the syntactic and the deep analysis, which is the most important and difficult level in Natural Language Processing. Logic-based approaches are rich research topics where there is still even though there is still room for improvement. This task was applied for many other languages (English, French, etc.) but not yet to Arabic. This is due to the lack of necessary tools of Arabic and the specificities of this language. Our approach is contribution for Arabic.

The collected questions refer to 115 factual questions. These can be than five categories, a PERSON: ؟ من صمم برج ايفل (*Who designed the Eiffel Tower?*), a LOCATION: ؟ أين تقع شلالات نياغرآ (*Where located the Niagara Falls?*), a DATE: متى استقلت تونس ؟ (*When Tunisia became independent?*), an ORGANIZATION: ماهي عاصمة ماليزيا ؟ (*What is the capital of Malaysia?*) or a NUMERIC EXPRESSION: كم يبلغ طول نهر الأمازون ؟ (*How much is the length of the Amazon River?*). Firstly, our questions analysis module extracts for each one's some necessary and relevant information (keywords, focus, type, etc.). In this regard, we implemented a tool that automatically interrogates the Web to extract the relevant passages that could answer these questions. Secondly, this analysis is accomplished by obtaining question transformation. This allows us to prepare the stadium to generate logical forms from the question. Indeed, these forms are used later in extracting the best answer.

This paper provides our study of analysing the question in order to generate the answer passage retrieval from the web. First of all, it begins with an introduction. Then, it suggests a typical question-answering architecture that steadily comprises three components (i.e. question analysis, document/passage retrieval and answer generation). Afterward, this paper discusses the question-answering in the Arabic language. In addition, this paper describes our proposed framework to generate Arabic passages of texts by querying the web. Once, collected questions are pre-treated, transformed in declarative form, and then into logic representation. Meanwhile, the passages of Arabic texts are recovered based on the elements obtained from the question analysis step. Finally, we discuss our corpus of pairs questions-texts called AQA-WebCorp.

## 2 Typical architecture of a question-answering system

A question-answering system corresponds generally to a processing chain bringing together three or four components that are more or less dependent. The techniques differ from one system to another; a typical architecture usually employs a pipeline architecture that chains together three main modules, namely: question analysis, document/passage retrieval and answer extraction. Each of these components deserves to be evaluated intrinsically, and also their assembly should be evaluated as a whole.

In this context, Bilotti and Nyberg [8] emphasize that proponents of the modular architecture naturally view the question-answering task as decomposable, and to a certain extent, it is. The modules, however, can never be fully decoupled, because question analysis and answer extraction components, at least, depend on a common representation for answers and perhaps also a common set of text processing tools. This dependency is necessary to enable the answer extraction mechanism to determine whether answers exist in the retrieved text, by analyzing it and comparing it against the question analysis module answer specification. In practice, the text retrieval component does not use the common representation for scoring text; either the question analysis module or an explicit query formulation component maps it into a representation queryable by the text retrieval component.

Hence, the pipelined modular question-answering system architecture also carries with it an assumption about the compositionality of the components. It is easy to observe that errors cascade as the question-answering process moves through downstream modules, and this leads to the intuition that maximizing performance of individual modules minimizes the error at each stage of the pipeline, which, in turn, should maximize overall end-to-end system accuracy.

- ✓ **Question analysis module:** This module takes as input a question in natural language and produces a set of question features, including, keywords, focus, expected answer type, etc. Depending on the retrieval and answer extraction strategies, some question analysis modules also perform syntactic and semantic analysis of the questions, such as dependency parsing and semantic role labeling. The main objective of this module is to obtain the features from the question that could be helpful in the following steps. All the information obtained by this module is given to the following steps of the system [9].

- ✓ **Document/passage retrieval module:** This module takes into account the information withheld during the analysis of the question. Generally it uses search engines to accomplish the recovery of documents or passages that could answer this question. Some question-answering systems use specific search engines like Indri [10] or Lucene. Some others query Google to ensure the passage's recovery.

✓ **Answer extraction module:** The answer extraction module identifies candidate answers from the relevant passage set and extracts the answer most likely to answer the user question [11].

It should be noted that there were other studies that also provide extra functionalities, such as, answer justification [12], query expansion using external resources (i.e., the Web) [13], [14]. Indeed, the answer justification module either takes the answer produced by the system and tries to verify it using resources such as the Web, or it uses external databases or other knowledge sources to generate the answer, and "project" it back into the collection to find the right documents. The query expansion is often performed since the questions can be quite short. Moreover, taking into account keywords from the question is not sufficient to provide contextual information for effective retrieval.

## 3 Arabic question-answering

In this section, we present a brief overview of the various tasks that are covered by several investigations carried out in Arabic, into the analysis of the question. This analysis varies from one study to another. More details are presented in Table 1 below. In this respect, most of these efforts focus on extracting the keywords or recognizing the named entities from the user question. In our proposal, the analysis of the question is determined essentially by defining the expected answer type, generating the focus of each question, extracting the keywords and transforming the question into declarative form then creating a logic representation.

*Table 1: Question analysis description: Arabic investigations*

| System | Tasks description |
|---|---|
| QARAB [15] | Extract type and category of desired answer (name, place, quantity…). |
| System of [16] | Extract question keywords; recognize question named entities, classify the question. |
| ARABIQA [17] | Classify the question; extract the keys words and the named entities |
| QASAL [18] | Formulate the query; extract the expected answer type, the question focus and the question key words. |
| System of [19] | Tokenize the question; determine the type and the focus (proper noun phrase), extract the root of all non-stop words. |
| DefArabicQA [20] | Identify the topic question (i.e., NE) and dedicate the expected type answer |
| AQuASys [21] | Identify the expected answer type; segment the question into interrogative noun, question's verb and question's keywords |
| IDRAAQ [13] | Extract the keys words; recognize the expected answer; create the query. |

| System of [22] | Remove the question mark and interrogative particle; tokenize; remove the stop words and the negation particles, tag, parse. |
|---|---|
| JAWEB [23] | Tokenize; detect the answer type; extract the question key words; generate the extra key word; stem the question key word. |
| Al-Bayan [24] | Classify the questions with Support Vector Machine; extract the question type, expected answer type and the named entities. |

Indeed, as shown in table 2 below, we find that the majority of Arabic systems, in Table 1 above, deal with question classification. In our case, the analysis of the question is often implemented in order to extract features, namely, keywords, focus, expected answer type and the declarative form of the question, which will be used not only for finding an accurate answer, but also for the treatment of other modules in the chain generation of this answer.

*Table 2: Question analysis tasks covered by Arabic investigations*

| | | Question processing tasks | | |
|---|---|---|---|---|
| | | Question Segmentation | Question Classification | Question Formulation |
| Question-answering | AQAS [25] | | | |
| | QARAB [15] | | ✓ | |
| | ArabiQA [17] | | ✓ | |
| | QASAL [18] | | | ✓ |
| | DefArabicQA [20] | | ✓ | |
| | AquASys [21] | ✓ | | |
| | IDRAAQ [13] | | ✓ | |
| | ALQASIM [26] | | | |
| | System of [22 | | | ✓ |
| | JAWEB [23] | | | |
| | Al-Bayan [24] | | ✓ | |

Referring to the literature of question-answering systems, the classification of question has been studied, among others, in several investigations other than Arabic. Moreale and Vargas-Vera [27] indicated that the question classification provides information about the kind of answer. The natural language question needs to be classified into various sets for extracting more precise sets of answers [28]. Some other recent works are focused on the question classification. These works attribute, to a given question written in natural language, one or more class labels depending on classification strategy [29]. Some other studies like [30], [31] and [32] emphasized that the performance of question classification has significant influence on the overall performance of a question-answering system.

Our approach is different from other ones carried out in Arabic question-answering systems in that we use the logic representation to analyze the Arabic statements and generate an accurate answer. In addition, we present in our survey [33] a performance analysis of the different investigations in Arabic. In fact, we explore an analysis of main question-answering tasks (question analysis, passage retrieval, and answer extraction). To analyze a given question, we suggest that most studies of Arabic question-answering, as is shown in table 2, are focused on question classification. However, in our research, the step of analyzing the question determines the expected answer type, the focus and the keywords of the question. And then, we propose to formulate and generate the declarative form of the question in order to generate the logic representation. This transformation could help us to extract an accurate answer.

## 4 A proposed method for question analysis

In this section, we will focus on how we analyze our questions. The challenge that this module is trying to respond is: What's the question about? In other words: What's the topic? What does the question mean? Indeed, we analyze the question to know what it means. The rest of this research paper explains how we analyze our collected question in order to generate their answers. As shown in figure 1 below, all features produced by this module are taken into account in the following steps of our Arabic question-answering system.

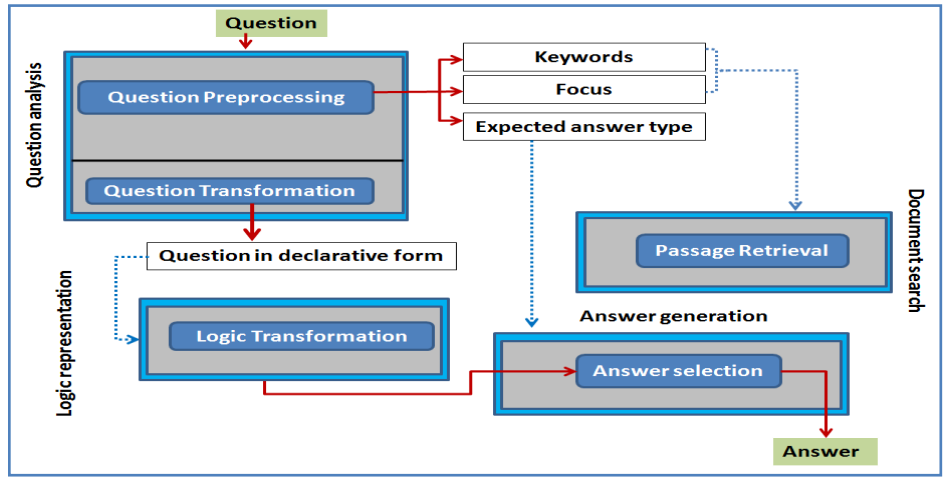

*Figure 1: Question analysis module*

### 4.1 Question collection and pre-processing

It is comes to collect a set of questions in natural language. These questions can be asked in different fields, including sport, history & Islam, discoveries & culture, world news, health & medicine. Our corpus consists of 115 questions and texts, which are extracted from the Web. Indeed, as shown in figure 2 below, the collection of these questions is carried out from multiple sources namely, discussion forums, frequently asked questions (FAQ), some questions translated from the two evaluation campaigns TREC and CLEF.

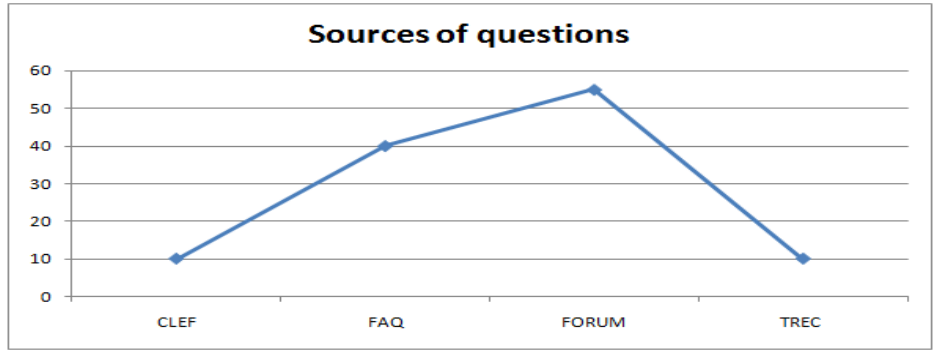

*Figure 2: Source of the questions used for our corpus*

The data collected from the web about the questions and the texts will help us to build an extensible corpus for Arabic question-answering. The size of our corpus is on the order of 115 factual questions: 10 questions translated from TREC, 10 questions translated from CLEF and 95 questions gathered from the forums and FAQs. To build our corpus, we used the Arabic texts available on the Internet that are collected on the basis of the questions posed at the outset.

After collecting such a set of questions, the search for the answer involves several steps, including, the analysis of the question, this is an equally important step, which is invaluable for identifying an accurate answer. The objective of this step is to obtain features from the question that could be helpful in the following steps. All information collected by this module is used in the following steps of our question-answering system. That is, we need to identify from each question a list of features that will then be used to search for relevant documents. Our question analysis module accomplishes two sub-steps. The first concerns the question pre-processing. The main objective of this sub-module is to extract the keywords from the question, to generate the focus and to derive the expected answer type from the question. This is a crucial step of the processing since the answer extraction module uses a strategy depending on the expected answer type. Then, the question type is typically related to the expected answer type, which in turn is typically related to the named-entity types available to the system. Particularly, our proposed question analysis module returns from some features (list of keywords and the focus) the relevant passages that contain those features.

Wang [34] noted that there is another source of information that is used by almost all question-answering systems is the named-entity list (NE). The idea is that factoid questions fall into several distinctive types, such as "location", "date", "person", etc. Assuming that we can recognize the question type correctly, then the potential answer candidates can be cut down to a few NE types that correspond to the question type. In order to determine the expected answer type, we propose to use the ArNER tool [35] to identify the type of the question from the named entities that contains. The different answer types that can be treated by our system are shown in Table 3 below.

*Table 3: Mapping question type to expected answer type*

| Type | | Example of questions (Arabic) | Example of questions (English) |
|---|---|---|---|
| Question type | Expected answer type | | |
| Who-من | Person | من صمم برج ايفل؟ | *Who designed the Eiffel Tower?* |
| Where-أين | Location | أين تقع شلالات نياغرا؟ | *Where located the Niagara Falls?* |
| When-متى | Date | متى استقلت تونس؟ | *When Tunisia became independent?* |
| What-ماهو/ماهي | Organization | ماهي عاصمة ماليزيا؟ | *What is the capital of Malaysia?* |
| How-كم | Numerical expression | كم يبلغ طول نهر الأمازون؟ | *How much is the length of the Amazon River?* |

In order to correctly answer a question, usually one needs to understand what the question asks for and to determine the relevant terms of the question. The definitions of these terms refer to the following example question "من صمم برج ايفل؟" (***Who designed the Eiffel Tower?***).

The keywords help the system to locate sentences where answers can probably be found [36]. The focus is the part of the question that is a reference to the answer [37]. In our case, the expected answer type refers to the named entities returned by ArNER, only paragraphs that contain a Name Entity of the same type that the expected answer type are validated. Hence, the named-entity answer extraction method selects any candidate answer that is an instance of the expected answer type [8]. In fact, almost all the Arabic question-answering systems involve keywords extraction. So, given some keywords it only returns the relevant passages that contain those keywords. Hence, it is necessary to analyze those passages to select an accurate answer. In the example above, if the system understands that the question asks for a person name, the search space of plausible answers will be significantly reduced. The focus is "برج ايفل" (*Eiffel Tower*), the keywords are "صمم, ايفل, برج" (*design, Eiffel, Tower*) and the expected answer type is "person".

**4.2 Question transformation**

We dedicate this section to obtaining the declarative form of each question. This transformation could be helpful in the logic representation module. The questions pre-processed and transformed in natural language are analyzed to get the information that can help us locate the correct answer.

As noted in Figure 1 above, our analysis of the question produces some elements that will be used later in other modules generating an accurate answer. For example, the list of keywords and the focus are used in searching the passages; the declarative form is used in the logic representation. The type of the expected answer is used in generating an accurate answer.

This current research describes the first step in our approach presented in [7] and [33], which is analyzing the question. A more detailed description of this module is shown here. Indeed, this module detects essentially the main features for each question, namely the list of question keywords, the focus of the question, and the expected answer type. With a real interrogation of Google, these characteristics may recover for each given question the extracts that address answers for this question. Then, this module infers the question in its declarative form in order to generate, as much as possible, logical representation for each question. Furthermore, the different modules of the extraction software identifying an accurate answer presented in our approach are strongly linked to the question analysis module. Firstly, the keywords and the focus are used to interrogate Google and retrieve relevant passages. Secondly, the declarative form is designed to infer the logic representation. After that, the expected answer type is carried out to select an accurate answer.

In summary, our study to analyze the question relies on Natural Language Processing tools. We performed experiments with various tools and we had to adapt some of them for achieving our aims. The expected answer type is generally identified by looking up the named entities of each question. To do this, we use the ArNER tool that was defined in the work of the team of Automatic Natural Language Processing of Mir@cl Laboratory [35]. Then, to transform the given question into its declarative form, we suppose to remove the interrogative particle and the question mark. At the end, to extract the set of keywords from our collected questions and to identify the focus of the question, we implement a script java. To generate the logic representation, we use the Al-khalil parser [38] to carry out the morphological analysis in order to identify the grammatical category of question words. In addition, segmentation and morphological analysis play a very important role in most applications of Natural Language Processing (i.e., information extraction, automatic summarization, etc.).

### 4.3 Logic representation

Many studies have investigated explicit logic forms and theorem proving techniques in question-answering, in other languages than Arabic. In this section, we present the background of the development of question-answering systems dealing with logic and inference approaches from their appearance to the present time. Most approaches adopt First Order Logic based formalisms.

First, in [39], the authors discuss a question-answering system that uses a theorem prover, based on Logic Form Transformation of question-answering. Therefore, both semantic transformations for questions and answers are translated into logic forms and presented to a simplified theorem prover.

Besides, Moldovan and Rus [40] discussed the conversion of WordNet glosses into axioms via LFT in the context of eXtended WordNet (XWN). In [41], the authors report on the implementation of the COGEX logic prover, which takes in question-answers LFs and WXN/NLP axioms and selects answers based on the proof score. In [42], the authors discuss enhancing the capabilities of COGEX by incorporating semantic and contextual information during LF generation.

As well, Mollá and his associates in [43] describe ExtrAns, a question-answering system applied to the Unix manual domain, which uses Minimal Logical Forms (MLFs) that are converted to Prolog facts/queries. In [44], the author compares MLFs with grammatical relations as the overlap-based similarity scoring measures for answer ranking.

After that, Rinaldi et al. [45] have explored a logic-based approach to biological question-answering, in adapting Molla et al.'s [43] ExtrAns system to the genomics domain. In adapting the ExtrAns system to the genomics domain, Rinaldi et al. worked with two domain-specific document collections: (1) GENIA corpus, and (2) 'Biovista' corpus consisting of full-text journal articles, generated from MEDLINE using two seed term lists concerning genes and pathways.

Furthermore, Benamara [46] developed a question-answering system applied to the tourism domain, called WEBCOOP, which contains facts, rules, and integrity constraints encoded in Prolog, and a set of texts indexed via FOL formulae.

Then, Clark and his associates [47] present a layered approach to the FOL representation of contextual knowledge, coupled with reasoning mechanisms, to enable contextual inference and default reasoning for question-answering. And then, Tari and Baral [48] proposed a question-answering system that uses AnsProlog for representation and reasoning.

However, Baral and his collaborators present [49] a question-answering system that combines AnsProlog and Constraint Logic Programming, to enable textual inference on events, actions, and temporal relations.

Finally, Terol et al. [50] have explored a logic-based approach, in adapting a generic restricted-domain question-answering system to the medical domain. The question-answering processing is based on the derivation of LFs (logic forms) from texts through the application of NLP techniques and on the complex treatment of the derived LFs.

Yet, in Arabic question-answering, there are some aspects that have been least researched. These aspects concern the use of semantic and the incorporation of logic and reasoning mechanisms. We have encountered only a few approaches that have been attempted in the semantic representation. Some approaches adapted question-answering approaches for making use of Arabic ontologies [3]. To the best of our knowledge, there are only a few studies that provide logic and inference based-approaches. For example, Bdour and Gharaibeh [22] proposed an Arabic question-answering system based on the paragraph retrieval, the authors used a corpus of 20 Arabic documents, and a collection of 100 different yes/no questions that are transformed into a logic representation. Nonetheless, with respect to their proposition, we don't find any information about sequencing it. So, a semantic logic based approach is essential and urgent. Furthermore, the lack or the absence of approaches of this kind in Arabic suggests the relevance and feasibility of exploring semantic logic-based approaches to Arabic question-answering.

In the rest of this section, we identify some rules for mapping Arabic statements into logic forms to be taken into consideration when looking at the use of textual entailment techniques that rely on both logic and semantic representation to extract the desired answer. These rules have been extracted and generalized from existing types of our collected questions. Generally, the logical representation attempts to capture the semantics (meaning) of the question [12]. Our work is concentrated on an implementation step to develop a question-answering system in Arabic using the techniques of textual entailment recognition. Text features extraction (keywords, named entities, relationships that link them) is considered the first step in our text modeling process. The second one is the use of textual entailment techniques that relies on logic and inference representation of Arabic statements to extract the candidate answer.

In order to determine which logic representation is best suited to a question of a specific type, we aim to propose some rules that can transform the Arabic statements into logic representation; more details are presented in Table 4 below. In addition, a predicate expression is a graph of predicate-argument relationship; we work with the following examples:

*Table 4: Logic representation rules for each question type*

| Question words | Logic Rule | Question Logic Representation (QLR) |
|---|---|---|
| **Question** | **من صمم برج ايفل ؟ (PERSON)** *(Who designed the Eiffel Tower?)* | |
| صمم *(design)* <br> برج *(Tower)* <br> ايفل *(Eiffel)* | PERSON (X) → من *(Who)* <br> VERB (X,Y) → صمم *(design)* <br> NOUN (Y) → برج *(Tower)* <br> NOUN (Y) → ايفل *(Eiffel)* | $\exists X, \exists Y, PERSON(X) \wedge$ صمم $(X,Y) \wedge$ برج $(Y)$ $\wedge$ ايفل $(Y)$ <br><br> $\exists X, \exists Y, PERSON(X) \wedge design(X,Y)$ $\wedge Tower(Y) \wedge Eiffel(Y)$ |
| **Question** | **أين تقع شلالات نياغرآ ؟ (LOCATION)** *(Where located the Niagara Falls?)* | |
| تقع *(locate)* <br> شلالات *(Falls)* <br> نياغرآ *(Niagara)* | LOCATION (X) → أين *(Where)* <br> VERB (X,Y) → وقع *(locate)* <br> NOUN (Y) → شلالات *(Falls)* <br> NOUN (Y) → نياغرآ *(Niagara)* | $\exists X, \exists Y, LOCATION(X) \wedge$ تقع $(Y,X) \wedge$ شلالات $(Y) \wedge$ نياغرآ $(Y)$ <br><br> $\exists X, \exists Y, LOCATION(X) \wedge$ تقع $(Y,X)$ $\wedge Falls(Y) \wedge Niagara(Y)$ |
| **Question** | **متى استقلت تونس ؟ (DATE)** *(When Tunisia became independent?)* | |
| استقل *(Became independent)* <br> تونس *(Tunisia)* | DATE (X) → متى *(When)* <br> VERB (X,Y) → استقل *(Became independent)* <br> NOUN (Y) → تونس *(Tunisia)* | $X, \exists Y, DATE(X) \wedge$ استقل $(Y,X) \wedge$ تونس $(Y)$ <br><br> $X, \exists Y, DATE(X) \wedge Became\ independent$ $(Y,X) \wedge (Tunisia)(Y)$ |
| **Question** | **ماهي عاصمة ماليزيا ؟ (ORGANISATION)** *(What is the capital of Malaysia?)* | |
| عاصمة *(Capital)* <br> ماليزيا *(Malaysia)* | ORGANIZATION (X) → ماهي *(What)* <br> NOUN (Y,X) → عاصمة *(Capital)* <br> NOUN (Y) → ماليزيا *(Malaysia)* | $X, \exists Y, ORANIZATION(X)$ $\wedge$ عاصمة $(Y,X) \wedge$ ماليزيا $(Y)$ <br><br> $X, \exists Y, ORANIZATION(X)$ $\wedge Capital(Y,X) \wedge Malaysia(Y)$ |
| **Question** | **كم يبلغ طول نهر الأمازون؟ (NUMERICAL EXPRESSION)** *(How much Reaches the length of the Amazon River?)* | |
| بلغ (Reach)<br>طول (Length)<br>نهر (River)<br>الأمازون (Amazon) | كم ← NUMERICALEXPRESSION (X)<br>(How much)<br>بلغ ← VERB (Y, Z, X) (Reach)<br>طول ← NOUN (Y) (Length)<br>نهر ← NOUN (Z) (River)<br>الأمازون ← NOUN (Z) (Amazon) | ∃ X, ∃ Y, NUMERICAL EXPRESSION<br>(Z)نهر∧(Y)طول∧(Y, Z, X)بلغ∧(X)<br>∧ الأمازون (Z)<br><br>∃ X, ∃ Y, NUMERICAL EXPRESSION<br>(Z)نهر∧(Y)طول∧(Y, Z, X)بلغ∧(X)<br>∧ الأمازون (Z) |

## 5  Answer Passage Retrieval

We conducted experiments on text passage extraction from the Web based on the elements retained in the question analysis. After collecting and analyzing the question, the following step is concerned with access to Google to retrieve passages that can answer the given question. Instead of returning a number of documents like Google, our passage retrieval module returns some passages or a few sentences of different lengths which can provide users with some contextual information for the answers.

As we can see, many of the features shown in our question analysis module appear in the above passage (ايفل, برج, صمم) *(design, Eiffel, Tower)*. This will allow the search module associated with our question-answering system to find an appropriate candidate passage. Thus, these features will also be used to extract the desired answer.

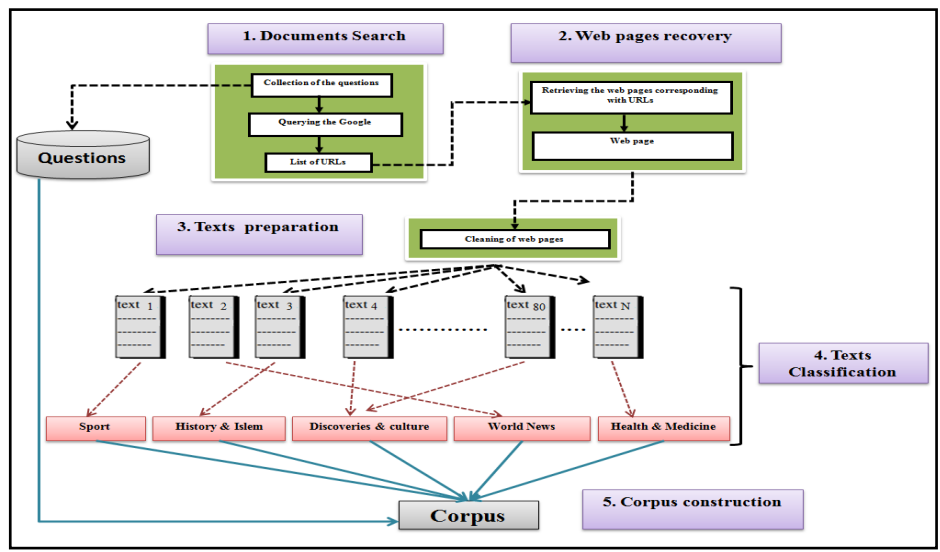

*Figure 3: Construction process of our corpus AQA-WebCorp*

This methodological framework is to look for web addresses corresponding to each question. Indeed, we have segmented questions in a list of keywords. Then, our tool seeks a list of URLs that match those keywords. Next, for each given address we propose to recover the webpage that it points to. In this respect, our corpus construction tool is an interface between the user request and Google. Specifically, it is a way to query the Google database to retrieve a list of documents. Finally, we performed a transformation of each retaining web page from ".html" format to ".txt" format. Finally, we look to see whether the answer is found in the corresponding text. The text is considered valid to build our body if it contains this answer. Otherwise, we go to the following URL (see figure 3 below).

### 5.1 Document retrieval

The idea consists in automatically generating the passages containing a particular word or in selecting only the passages containing variations of the words of the question. In our case, to find an answer to a question in Arabic, we propose to use a search engine (i.e., Google) to retrieve the documents related to each question. Then, we add post linguistic treatments to those documents that actually constitute our corpus to select an accurate and appropriate answer. In this respect, querying a search engine accelerates the recovery of documents online but requires offline processing of these documents. At this stage, the document search module has been implemented. First, when a question is asked, our tool submits it to the search engine (Google) to identify the list of URLs based on a list of the words that constitute this question.

### 5.2 Passage generation

Let's take the following example: our tool can then, from the question: « من صمم برج ايفل ؟ » **(Who designed the Eiffel Tower?),** generate a list of equivalent URLs. In addition, Google is the default access means is through a search engine. By clicking the "search URLS" button, a list of addresses automatically exposed. And then, for each URL, this prototype can retrieve the necessary information (host, query, protocol, etc.). When the list of URLs is determined, our tool extracts for each address the corresponding web page. For each given URL, we propose to find the corresponding HTML page; figure 4 illustrates this case. From the address retained in the first step, a set of web pages is recovered. Each web page is exported in ".html" format.

Thus, we propose to transform the HTML web page obtained in the previous step into a ".txt" format. The texts are being in ".html" format, it seems justified to put them in the ".txt" format. On the basis of the question cited above, let us consider the extract of the following relevant passages which may contain the correct answer, as shown in figure 4 below. For this, we remove all the HTML tags for each retrieved page. It is possible to either keep the text for our own corpus construction work, or discard it.

```
Passage 1 :
...مصمم تمثال الحرية و برج إيفل هو نفسه المهندس الذي صمم كوبري إمبابة بالقاهرة، و هو المعماري الفرنسي جوستاف إيفل.
Passage 2 :
...الموقع الرسمي . وثيقة حياة غوستاف إيفل - موقع برج إيفل - موقع حياة غوستاف إيفل - وثيقة حياة غوستاف إيفل (بالفرنسية)؛ وثيقة حياة غوستاف إيفل (بالإنكلزية)؛ موقع برج إيفل - موقع حياة غوستاف إيفل.. &بالفرنسية Alexandre Gustave)) (ألكسندر غوستاف إيفل (لقب عائلته بونيكهوسن,...
Passage 3 :
...أب (أغسطس) 2015 ... ولكن عبقرية جوستاف إيفل لم تتوقف على بناء برج إيفل فقط بل هو قام ... انه تمثال يشير الى الحرية حيث انه صمم على شكل امرأة في يدها اليمنى تحمل &29
Passage 4 :
...تشرين الثاني (نوفمبر) 2010 ... بنى برج إيفل امام قوس النصر في فرنسا بهدف مناسبة مرور مئة عام على الثورة الفرنسية عام 1889 قصة برج إيفل كاملة مع الصور لك ارث مصمم البرج &11
Passage 5 :
...جوستاف إيفل وهو مهندس برج إيفل كان المتوقع أن يظل البرج لمدة 20 سنة فقط ولكن تم استعماله في الاتصالات ومن ثم الحرب وكان ينظر اليه كرمز للنصر الحديد المستعمل&
Passage 6 :
;نيسان (إبريل) 2014 ... محتويات &1633#؛ برج إيفل &1634#؛ من بنى برج إيفل؟ &1635#؛ بناء البرج &1636#؛ وصف البرج &1637#؛ زيارة برج إيفل برج إيفل الفرنسي الرائع والصرح الشاهق الذي 16
Passage 7 :
...حقائق مذهلة عن برج إيفل بعد رمز العاصمة الفرنسية و من أشهر المزارات السياحية على مستوى العالم و يتوافد السياح من جميع بقاع الأرض لزيارته والتقاط&10
```

*Figure 4: Example of a text containing an answer to the following question:* " من صمم برج ايفل ؟ " *(Who designed the Eiffel Tower?)*

## 6 AQA-WebCorp: an experimental Corpus Preparation

With a corpus, qualitative and quantitative linguistic research can be done in seconds, saving time and effort. Initially, we identified several elements of the analysis of the questions that can facilitate the generation of answers. Finally, empirical data analysis can help researchers not only to conduct effective new linguistic research, but also to test existing theories. We used Google to search for documents containing at least an answer to questions. This is to make the most of the most likely text passages that contain the answer to a given question. Indeed, the search for these passages is done using the Web as a resource and collection of documents and relying on the important criteria extracted from the questions.

The corpus construction task from the Web was discussed for different applications of Automatic Natural Language Processing, not only for the question-answering task. The performance of a question-answering system is highly dependent on a good source corpus and accordingly well formalized users' demands. If the corpus is structured and users' demands are well formalized, then the burden is on the question-answering system to use complex Natural Language Processing techniques to understand how the text is reduced [51].

Our corpus consists of texts come from four sources, namely TREC, CLEF, discussion forums and frequently asked questions (FAQ). TREC and CLEF are two of major information retrieval evaluation forums in the world. Their evaluation tracks are in Natural Language Processing and Information Retrieval domains such as large-scale information retrieval, question-answering, cross language processing, and many new hot research topics. Furthermore, we are currently developing a corpus dedicated to Arabic question-answering. The size of the corpus is on the order of 115 pairs of questions and texts. This was collected using the web as a source of data. The data collected, of the questions and the texts from the web, will help us to construct a corpus for Arabic question-answering. The pairs of texts-questions are distributed in five areas " أخبار العالم; التاريخ والإسلام; إكتشافات وثقافة; رياضة; صحة و طب" (world news, history & Islam, discoveries & culture, sport, health & medicine) as shown in figure 5 below:

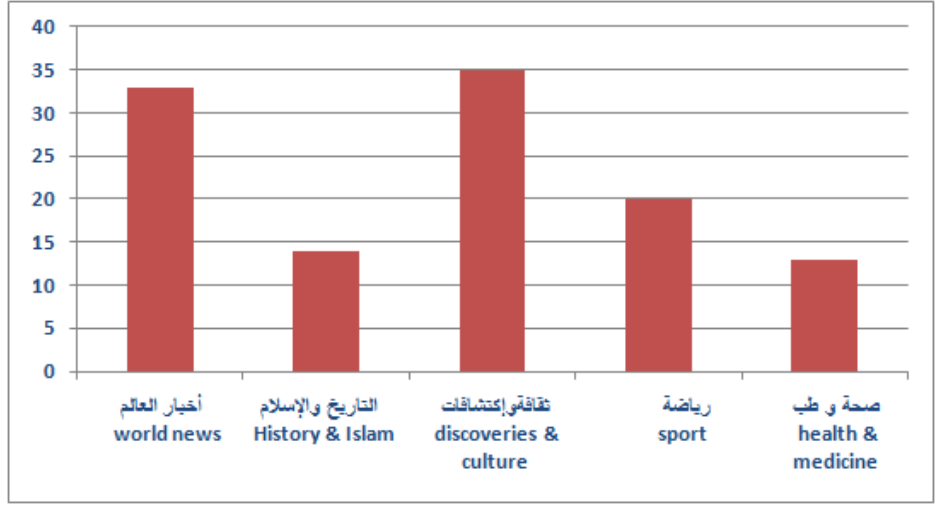

*Figure 5: Statistics of question-text pairs used in different sources*

To implement our corpus for Arabic, we propose a simple and robust method implemented in Java. The principle of this method is based on four stages, relatively dependent. The construction of our corpus of pairs of Arabic question and texts is actually done by developing all of these four steps. We implemented two modules; the first automatically analyzes the question by offering some features of each question, like focus, expected answer type, keywords, etc. Then, the second extracts passages that could contain the answers of these questions.

In this paper, we have followed a theoretical method validated by an empirical investigation to provide better analysis and understanding of each question. This initiative eventually produces a declarative form for all questions followed by a logical representation. It is therefore our goal to provide a better platform to retrieve passages that can answer these questions. This can then achieve a better understanding of those passages to know what is there in the text collection and to select the correct answer to each question.

### 6.1 Question collection and analysis

In all question-answering systems, the generation of a precise answer to a natural language question necessarily involves a step of analyzing this question. From our 115 collected questions, we selected five different question types: expected answer type, person, location, temporal expression, organization, numerical expression. As shown in figure 6 below, for each question type, we extracted the corresponding features: the focus, the expected answer type and the list of keywords. Next, we translate the question into its declarative form. Then, we transform it in a logic representation. Our question analysis module assumes each question to be a simple declarative sentence, which is composed of a sequence of words and searches for a focus for each sentence as useful evidence to extract an accurate answer. We describe the process of analysis with examples of 5 types of questions collected in our corpus. This procedure is repeated for different examples of the same question type.

In summary, our contribution is to design and implement a prototype for analyzing Arabic factual questions. The aim is to present some features from each question. Those features can aid us later to select the relevant passages from the web. To do it, we implement a Java script that analyzes our collected questions and interrogates the web to research the relevant passages in which an accurate answer is located.

Consider the following example of a question, for which the type of expected answer is a person. Analysis of the question gives rise to the following features:

- ✓ Keywords: صمم , برج, ايفل  *(design, Eiffel, Tower)*
- ✓ Focus: برج ايفل *(Eiffel Tower)*
- ✓ Expected answer type: person
- ✓ Declarative form : صمم برج ايفل *(designed the Eiffel Tower)*
- ✓ Logic representation : ∃ X, ∃ Y, PERSON (X)∧ صمم (X,Y)∧ برج (Y)∧ ايفل (Y) *(∃ X, ∃ Y, PERSON (X) ∧ design(X,Y) ∧ Tower(Y) ∧ Eiffel(Y))*

Our two proposed modules, which concerned question analysis and passage retrieval, are implemented in Java. The snapshot of these components has been shown below.

**6.2    Answer passage retrieval**

Passage retrieval is a way to query the Web to retrieve a passage that answers a given question. The passage recovery module returns some passages or a few sentences of different lengths that can provide users with contextual information for the answers. The first stage starts with a question posed in natural language, analyzes the question, and produces a list of corresponding URLs that may contain the candidate answer. In addition, the method for finding the relevant passage is described in [52]. Indeed, the document search is based on the words of the collected questions. The better to ensure this step we developed a Java script for interrogating Google to obtain these results. The result of this step is a set of URLs addresses. For each question a list of the URLs will be constructed.

**6.3    Results**

The evaluation is an essential step in the development of a computer application for the NLP, and especially the question-answering systems. The evaluation of a question-answering system can be done for the whole system and/or for each module, especially the passage retrieval module and the answer validation module. The proposed method is effective despite its simplicity. We managed to demonstrate that the Web could be used as a data source to build our corpus. In this section, we present empirical evaluation results to assess the number of questions that are correctly analyzed and translated into the logic form. Typically, the performance of a question transformed into the logic form is measured by calculating the accuracy. The measure in question logic translation is defined as follow:

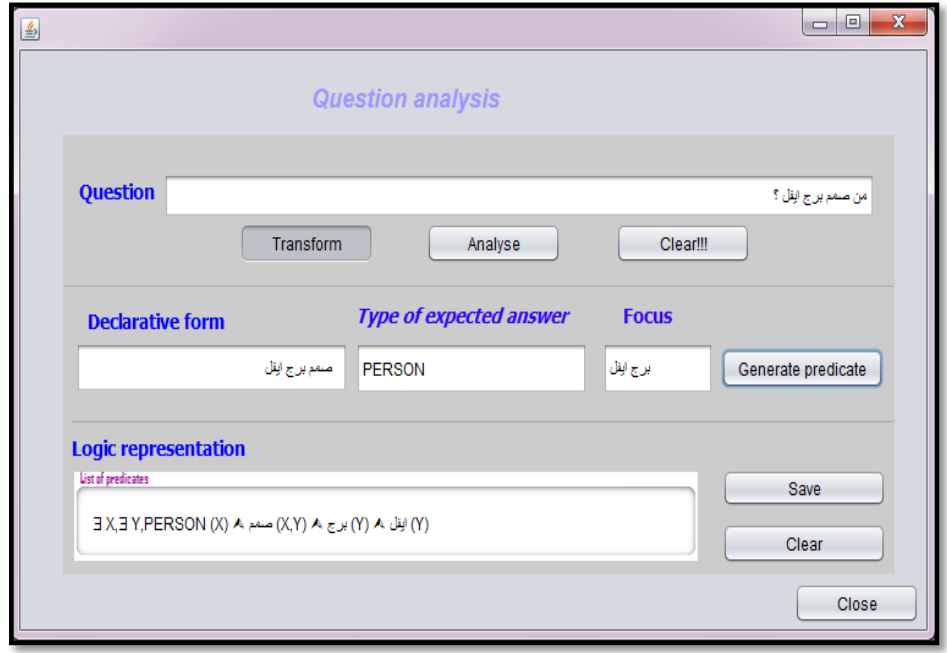

*Figure 6: Proposed tool for question analysis*

We included experiments on question analysis and answer passage retrieval. First, note that the evaluation of our method for anlyzing the question focused on a collection consists of 115 factual questions collected from four different sources, 10 questions translated from TREC, 10 questions translated from CLEF and 95 questions gathered from the forums and FAQs. We obtained 74 correct logic transformations of the analysed 115 questions and 41 false ones (15 false transformation are mainly due to errors in named entity recognition, 11 are mainly due to errors on question transformation in the declarative form and 15 of the 115 questions, we were not able to identify the logic representation. The evaluation gives an Accuracy value of 0.64.

The Accuracy measures the number of questions correctly transformed divided by the total number of collected questions (correctly translated and false translated).

$$\textbf{Accuracy } = \frac{\textbf{CT}}{\textbf{TQ}} \qquad (1)$$

Second, we tested our proposed method for Answer passage retrieval with a set of collected questions, which consists of 115 factual questions. To evaluate our method, two performance measures are employed, Accuracy and c@1. These measures are used to measure the passages of texts automatically generated from the web that could answer those questions. In our case, there are some questions that are not answered or incorrectly answered. According to [53], not answering has more value than answering incorrectly. To evaluate those unanswered questions, we use the c@1 measure. It is an extension of the accuracy measure (the proportion of correctly answered questions). This measure has a good balance of discrimination power, stability and sensitivity properties.

**CA**: Number of Correct Answers.
**UQ**: Number of Unanswered Questions.
**TQ**: Total Number of Questions.

The Accuracy measures the number of questions correctly answered divided by the total number of collected questions (correctly answered and not correctly answered).

$$\textbf{Accuracy } = \frac{CA}{TQ} \quad (2)$$

The c@1 measures the proportion of correctly answered questions.

$$\textbf{C@1 } = \frac{(CA\ +\ UQ\ *\ (CA\ /\ TQ))}{TQ} \quad (3)$$

It should be noted that the number of questions correctly answered is 101, and the number of questions which are incorrectly answered or not answered is 14 questions. So we get an Accuracy of 0.87 and a c@1 of 0.9. We also give the performance for the questions that are correctly analyzed and translated into the logic form; the results of our experiments are presented in Table 5 below.

*Table 5: Results of preliminary experiments*

|  | Question analysis | | | Answer passage retrieval | | | | |
|---|---|---|---|---|---|---|---|---|
|  | TQ | CT | Accuracy | Correctly answered | Unanswered = (**incorrectly** +**not** (answered)) | Total | Accuracy | C@1 |
| **Number** | 115 | 74 | 0.64 | 101 | 14 | 115 | 0.87 | 0.98 |

## 7 Conclusion and perspectives

In our research, we present an approach that analyzes the question, retrieves the text containing the answer and analyzes it. It transforms the question and the text into logical representations and extracts an accurate answer. Then, we recognize all entailments between them. The results of this recognition process are a set of text sentences that can imply the declarative form of the user's question. There are a number of directions for future research with a system that can produce logic representations for Arabic sentences that are robust enough to be used in making correct inferences. Essentially, our work is concentrated on the development of a question-answering system in Arabic using the techniques of textual entailment recognition. The extraction of text features (keywords, named entities, relationships that link them) is actually considered the first step in our text modelling process. The second one is the use of textual entailment techniques that rely on inference and logic representation to extract a candidate answer.

It would appear that the integration of logic-based approaches in Arabic question-answering systems can provide fairly good results. With our research results, we can observe how this kind of searching might be integrated in Arabic. There are some investigations for this task with its experiments in other languages, but through our interactive work we think we have some possible strategies for improving Arabic question-answering by proposing a new based logic approach for this language; and then, by experimenting with the phases of this approach. The proposed method is effective despite its simplicity. We managed to demonstrate that the Web could be used as a data source to build our corpus. The Web is the largest repository of existing electronic documents. Indeed, as prospects in this work, we have labelled this vast corpus and make it public and usable to improve the automatic processing of Arabic.

Our study attempts also to answer a series of questions including, how to analyze text passages generated from the web and how to select an accurate answer.

**Acknowledgements**

I give my sincere thanks for my collaborators Professor Patrice BELLOT (University of Aix Marseille, France) and Mr Mahmoud NEJI (University of Sfax-Tunisia) that i have benefited greatly by working with them.

# References

[1] Sinclair J (2005) Corpus and text - basic principles In: Wynne M (ed) Developing Linguistic corpora: A guide to good practice, Oxbow Books, Oxford, UK pp 1-16.

[2] Rastier F (1998) Enjeux èpistèmologiques de la linguistique de corpus, In: Williams CG (ed) La linguistique de corpus, Presses Universitaires de Rennes, Rennes, France, pp 31-46.

[3] AlAgha I, Abu-Taha A (2015) AR2SPARQL: An Arabic Natural Language Interface for the Semantic Web, International Journal of Computer Applications, 125(6), pp 19-27.

[4] Al-Khalifa H, Al-Wabil A (2007) The Arabic language and the semantic web: Challenges And opportunities, In: The first international symposium on computers and the Arabic language, November 2007, Riyadh, Saudi Arabia, pp 27-35.

[5] Meftouh K, Smaili K, Laskri MT (2007) Constitution d'un corpus de la langue arabe à partir du Web, Iera, Rabat, Morocco, 17-18, juin 2007. HAL Id: inria-00186536, version 1.

[6] Resnik P (1998) Parallel stands: A preliminary investigation into mining the web for bilingual text, In: Conference of the Association for Machine Translation in the Americas, pp 72-82, Springer Berlin Heidelberg.

[7] Bakari W, Trigui O, Neji M (2014) Logic-based approach for improving Arabic question-answering, In: IEEE international conference on computational intelligence and computing research (ICCIC), University of Tlemcen, Algeria, pp 1-6.

[8] Bilotti MW, Nyberg E (2008) Improving text retrieval, precision, and answer accuracy in question answering systems. In: Coling 2008, Proceedings of the 2nd workshop on information retrieval for question answering systems, Association for Computational Linguistics, Stroudsburg, PA, pp 1-8.

[9] Rodrigo Á, Perez-Iglesias J, Peñas A, Garrido G, Araujo L (2010) A Question Answering System based on Information Retrieval and Validation, In: CLEF (Notebook Papers/LABs/Workshops).

[10] Metzler D, Croft B (2004) Combining the language model and inference network Information approaches to retrieval. In: Information Processing and Management, Special Issue on Bayesian Networks and Information Retrieval, 40(5), pp 735-750.

[11] Mervin R (2013) An overview of question answering system, International Journal of Research in Advanced Technology (IJRATE), (1).

[12] Nyberg E, Mitamura T, Callan J, Carbonell J, Frederking R, Collins-Thompson K (2003) The JAVELIN question-answering system at TREC 2003: A multi-strategy approach with dynamic planning.

[13] Abouenour L, Bouzoubaa K, Rosso P (2012) IDRAAQ: New Arabic question answering system based on query expansion and passage retrieval, In: CLEF (Online working notes/labs/workshop).

[14] Ganesh S, Varma V (2009) Exploiting structure and content of Wikipedia for query expansion in the context of question answering, In: Proceedings of the international conference on recent advances in natural language processing (RANALP), pp 103-106.

[15] Hammo B, Abuleil S, Lytinen S, Evens M (2004) Experimenting with a question Answering system for the Arabic language, In: Computers and the Humanities, 38(4), pp 395-415.

[16] Rosso P, Benajiba Y, Lyhyaoui A (2006) Towards an Arabic question-answering system, In: Proceedings of the 4th scientific research outlook and technology development in the Arab world (SROIV) Damascus, Syria, 11-14 December, pp 11-14.

[17] Benajiba Y, Rosso P, Lyhyaoui A (2007) Implementation of the ArabiQA question answering system's components, In: Proc. workshop on Arabic natural language processing, 2nd Information Communication Technologies Int. Symposium. ICTIS-2007, Fez, Morocco, April, pp 3-5.

[18] Brini W, Ellouze M, Mesfar S, Belguith LH, (2009) An Arabic question answering system For factoid questions, In: Proceedings of the IEEE international conference on natural language processing and knowledge engineering (IEEE NLP-KE'09), Dalian, China, pp 1-7.

[19] Kanaan G, Hammouri A, Al-Shalabi R, Swalha M (2009) Question answering system for the Arabic language, American Journal of Applied Sciences, 6(4), pp 797-805.

[20] Trigui O, Belguith LH, Rosso P (2010) DefArabicQA: Arabic definition question answering system. In: Workshop on language technologies for Semitic languages, 7th LREC, Valetta, Malta, pp 40-45.

[21] Bekhti S, Al-Harbi L (2013) AQuASys: A question-answering system for Arabic, In: WSEAS International Conference. Proceedings. Recent Advances in Computer Engineering Series, 25(6), pp 19-27.

[22] Bdour WN, Gharaibeh, NK (2013) Development of yes/no Arabic question answering system, In International Journal of Artificial Intelligence & Applications (IJAIA). Vol.4. No.1, pp 51-63, January 2013. DOI: 10.5121/ijaia.2013.4105.

[23] Kurdi H, Alkhaider S, Alfaif N (2014) Development and evaluation of a web based question answering system for Arabic language, In: Computer Science & Information Technology (CS&IT) 4(2), pp 202-214.

[24] Abdelnasser H, Mohamed R, Ragab M, Farouk B (2014) Al-Bayan: An Arabic question-answering system for the Holy Quran, In: Proceedings of the workshop on natural language processing, pp. 57-64, Doha, Qatar, 10/25/14, Association for Computational Linguistics, Stroudsburg, PA, USA.

[25] Mohammed FK, Nasser K, Harb H. (1993) A knowledge based Arabic question answering system (AQAS). ACM SIGART Bulletin. 4(4), pp 21-30.

[26] Ezzeldin AM, Kholief MH, El-Sonbaty, Y (2013) ALQASIM: Arabic language question answer: selection in machines, In the 13th International Arab Conference on Information Technology, CLEF'2013, Springer Berlin Heidelberg, pp 100-103.

[27] Moreale E, Vargas-Vera M (2004) A question-answering system using argumentation, In: Proceedings of MICAI 2004, Advances in artificial intelligence, pp 400-409, Springer Publishers, Berlin, Heidelberg.

[28] Athira PM, Sreeja M, Reghuraj PC (2013) Architecture of an ontology- based domain specific natural language question answering system, International Journal of Web & Semantic Technology, IJWesT, 4, pp. 31-39.

[29] Loni B, Van Tulder G, Wiggers P, Tax DM, Loog M (2011) Question classification by weighted combination of lexical, syntactic, and semantic features, In: Proceedings of the conference on text, speech, and dialogue, pp 243-250, Springer Publishers, Berlin, Heidelberg.

[30] Ittycheriah A, Martin F, Zhu WJ, Adwait R, Mammone RJ, IBM's statistical question answering system, In: Proceedings of the 9th text retrieval conference, pp 229-237, NIST, Gaithersburg, MD, USA.

[31] Hovy E, Geber L, Hermjakob, U, Lin CY, Ravichandran D (2001) Toward semantics-based answer pinpointing, Proceedings of the first international conference on human language technology research, pp 1-7. March.

[32] Moldovan D, Pasca M, Harabagiu S, Surdeanu M (2003) Performance issues and error analysis in an open-domain question answering system, ACM Transactions On Information Systems, 21(2), pp 133-154, April, 2003.

[33] Bakari W, Bellot P, Neji M, (2015) Literature review of Arabic question-answering: In: Proceedings of the Eleventh Flexible Query Answering Systems 2015 (FQAS-2015), T Andreasen et al. (eds.), pp 321-334, Advances in Intelligent Systems and Computing, #400, Springer Publishers, Berlin, Heidelberg.

[34] Wang M (2006) A survey of answer extraction techniques in factoid question answering, HLT-NAACL, pp 1-13.

[35] Zribi I, Hammami SM, Belguith LH (2010) L'apport d'une approche hybride pour la reconnaissance des entités nommées en langue arabe, In: TALN'2010, pp19-23.

[36] Vicedo JL, Ferrandez A (2001) A semantic approach to Question answering systems, *NIST SPECIAL PUBLICATION SP*, (249), pp 511-516.

[37] Lally A, Prager JM, McCord MC, Boguraev BK, Patwardhan S, Fan J, Chu-Carroll J (2012) Question analysis: How Watson reads a clue, IBM Journal of Research and Development, 56(3/4)2, pp1-14.

[38] Bebah M O A O, Meziane A, Mazroui A, Lakhouaja A (2011) Alkhalil morpho sys, In 7th International computing conference in Arabic.

[39] Harabagiu SM, Pas¸ca, M, Maiorano SJ (2000) Experiments with open-domain textual question-answering, In: Proceedings of the 16th international conference on computational linguistics (COLING 2000), Saarbrucken, Germany, pp 292-298.

[40] Moldovan D, Rus V (2001) Logic form transformation of WordNet and its applicability to question answering. In: Proceedings of the Association for Computational Linguistics, (ACL'2001), pp 402-409, Toulouse, France.

[41] Moldovan D, Clark C, Harabagiu S, Maiorano SJ (2003): COGEX: A logic prover for question answering, In Proceedings of the 2003 Conference of the North American Chapter of the Association for Computational Linguistics on Human Language Technology-Volume , pp 87-93, Association for Computational Linguistics.

[42] Moldovan D, Clark C, Harabagiu S, Hodges D (2007) Cogex: A semantically and contextually enriched logic prover for question answering, J. App Logic, 5, pp 49-69, https://utdallas.influuent.utsystem.edu/.../cogex-a-semantically-and-contextually-enriched.

[43] Molla D, Schwitter R, Hess M, Fournier R (2000) ExtrAns, an answer extraction system, TAL. Traitement automatique des langues, 41(2), pp 495-522.

[44] Molla D, (2003) Towards semantic-based overlap measures for question-answering, In:Proceedings of the first Australasian language technology workshop (ALTW'03), pp1-8.

[45] Rinaldi F, Downdall J, Schneider G, Persidis A (2004) Answering questions in the genomics domain, In: Proceedings of the ACL 2004 workshop on question answering in restricted domains, pp 46-53, In: Proceedings of the ACL-2004 workshop: question answering in restricted domains, pp 31-38, Barcelona, Spain, Association for Computational Linguistics, Stroudsburg, PA, USA.

[46] Benamara F (2004) Cooperative question-answering in restricted domains: the WEB-COOP experiment In: Proceedings of the ACL-2004 workshop: question answering in restricted domains, pp 31-38, Barcelona, Spain, Association for Computational Linguistics, Stroudsburg, PA, USA.

[47] Clark C, Hodges D, Stephan J, Moldovan D (2005) Moving QA Towards reading comprehension using context and default reasoning, In: Proceedings of the AAAI 2005 workshop on inference for textual question answering, pp 6-12, Pittsburgh, PA, USA, AAAI Press, Palo Alto, CA, USA.

[48] Tari L, Baral C (2005) Using AnsProlog with Link Grammar and WordNet for QA with deep reasoning, In : Information Technology, 2006. ICIT'06. 9th International Conference on. IEEE, 2006. pp 125-128, Pittsburgh, AAAI Press Palo Alto, CA, USA.

[49] Baral C, Gelfond G, Gelfond M, Scherl RB (2005) Textual inference by combining multiple logic-forming paradigms, In: Proceedings of the AAAI 2005 workshop on inference for textual question answering, pp 1-5.

[50] Terol RM, Martinez-Barco P, Palomar M (2007) A knowledge-based method for the medical question-answering problem, Computers in Biology and Medicine, 37(10), pp 1511-1521.

[51] Mishra A, Jain SK (2015) A survey on question answering systems with classification, Journal of King Saud University - Computer and Information Sciences, 28(3), pp 345-361.

[52] Bakari W, Bellot P, Neji M, (2016) AQA-WebCorp: Web-based Factual Questions for Arabic, In: Proceedings of the 20th International Conference on Knowledge-Based and Intelligent Information & Engineering Systems (KES-20156), Procedia Computer Science, 96, pp 275-284.

[53] Peñas A, Rodrigo LA (2011) A simple measure to assess non-response, In: Proceedings of the 49th Annual Meeting of the Association for Computational Linguistics: Human Language Technologies, 1, pp 1415-1424.